\documentclass[aps,prb,reprint,longbibliography]{revtex4-2}
\newcommand{\degree}{^\circ}
\usepackage{amsmath}
\usepackage{amssymb}
\usepackage{graphicx}
\begin{document}
\title{Quasi-one-dimensional sliding ferroelectricity in NbI$_4$} 
\author{Ning Ding}
\author{Haoshen Ye}
\author{Shuai Dong}
\email{Email: sdong@seu.edu.cn}
\affiliation{Key Laboratory of Quantum Materials and Devices of Ministry of Education, School of Physics, Southeast University, Nanjing 21189, China}
\date{\today}
\begin{abstract}
Sliding ferroelectricity was originally proposed to elucidate the out-of-plane polarization generated by a specific stacking arrangement of non-polar van der Waals layers. However, the concept of sliding ferroelectricity can be generalized to more geometries. Here, the NbI$_4$ bulk is theoretical demonstrated as a quasi-one-dimensional sliding ferroelectric material, which exhibits a polarization of $0.11$ $\mu$C/cm$^2$ perpendicular to the Nb's chains. The most possible ferroelectric switching path is found to be via the interchain sliding along the chain direction, while other paths like Peierls-dimerization of Nb pairs may also work. Moreover, its polarization can be augmented for $82\%$ by hydrostatic pressure up to $10$ GPa, beyond which NbI$_4$ becomes a polar metal. In addition, the negative longitudinal piezoelectricity is also predicted.
\end{abstract}
\maketitle
	
\section{Introduction}
In recent years, low-dimensional ferroelectrics have emerged as a promising branch of polar materials, which may overcome the critical size effect to maintain stable ferroelectric polarization in the atomic limit \cite{wang2023-NM,zhang2023-NRM,Wumenghao-wires-2018,guan-AEM-2020,qi2021-AM}. Among two-dimensional (2D) ferroelectrics, there are several extensively-studied sub-branches including CuInP$_2$S$_6$ \cite{Liu-Nat-Com-2016}, $\alpha$-In$_2$Se$_3$ \cite{Ding-Nat-Com-2017}, SnTe \cite{chang2016-Science}, NbOI$_2$ \cite{jia-NH-2019}, and WO$_2$Cl$_2$ \cite{Lin-PRL-2019}. All of them belong to the ion displacement type ferroelectrics.
	
Interlayer sliding ferroelectricity, proposed by Wu \textit{et al} \cite{li-ACS-Nano-2017}, widely exists in 2D van der Waals (vdW) materials. Their out-of-plane ferroelectric polarizations are induced by specific stacking modes, which can be reversed by interlayer sliding \cite{wu2021-PNAS}. Till now, the experimentally confirmed sliding ferroelectrics include WTe$_2$ bilayer/few-layer/flakes \cite{Fei-Nature-2018,sharma2019-SA,xiao2020-NP},  hexagonal BN bilayer \cite{yasuda2021-Science,vizner2021-Science}, $R$-$MX_2$ ($M$=Mo/W, $X$=S/Se) \cite{wang2022-NN}, 1T'-ReS$_2$ bilayer \cite{wan2022-PRL}, and InSe bilayer \cite{sui2023-NC}. Besides these binary-element compounds, the sliding ferroelectricity was also reported in inorganic-organic hybrid vdW crystal (15-crown-5)Cd$_3$Cl$_6$ \cite{miao2022-NM}. In addition, a large number of sliding ferroelectrics have been theoretically predicted \cite{liu2020-PRL,zhang2021-PRB, ma2021-npjCM, ding2021-PRM, chen2024-NL, meng2022-NC, liu2023-npjCM,yang2023-PRL, liang2021-npjCM, xiao2022-npjCM, zhou2022-npjMA}. Furthermore, a general theory for bilayer stacking ferroelectricity  was also established based on the group theory analysis \cite{ji2023-PRL}, which can provide the design rule for slide ferroelectric bilayers. 
	
The ferroelectricity in quasi-one-dimensional systems has also received lots of attention, such as PVDF-TrFE nanowire \cite{hu2009-NM}, group-IV metal chalcogenides nanowires and nanotubes \cite{zhang2019-JACS, song2024-JMCC}, SbN/BiN nanowires \cite{yang2021-ACSAMI}, vdW oxyhalides WO$X_4$ and NbO$X_3$  ($X$=Cl, Br, I) \cite{lin2019-PRM,sun2022-PRM,du2022-npjCM}. Because of the quasi-one-dimensional characteristics, the theoretical upper-limit for ferroelectric memory density can reach $\sim100$ Tbits/in$^2$ \cite{lin2019-PRM}. Therefore, it will be interesting to generalize the concept of sliding ferroelectricity into the quasi-one-dimensional systems, because the sliding ferroelectricity is naturally superior to other non-sliding ferroelectrics regarding the  switching energy \cite{he2024-AtM}. However, the sliding ferroelectricity has not been touched in quasi-one-dimensional systems.
	
In this Letter, the quasi-one-dimensional NbI$_4$ vdW bulk with $4d^1$ electron configuration will be theoretically studied. The two chains within one primitive cell can slide, which lead to a dipole perpendicular to the chain direction. Nontrivially, this interchain sliding is driving by the intrachain Peierls dimerization. Multiple ferroelectric switching paths are allowed to reverse the polarization. Furthermore, its sliding polarization can be enhanced by $82$\% under $10$ GPa hydrostatic pressure, beyond which it becomes a polar metal. The intriguing negative longitudinal piezoelectric effect is also revealed in this vdW material.
	
\section{Methods}
The first-principles calculations were performed with the projector augmented-wave (PAW) potentials as implemented in the Vienna {\it ab initio} Simulation Package (VASP) \cite{PAW-PRB-1996}. The Perdew-Burke-Ernzerhof (PBE) parameterization of the generalized gradient approximation (GGA) was used for the exchange-correlation functional \cite{PBE-PRL-1996}. The plane-wave cutoff energy was $500$ eV, and the Nb's $5s4d4p4s$ electrons were treated as valence states. The $k$-point grid of $5 \times 5 \times 3$ was employed for structural relaxation and static computation. To describe the interchain interaction, two different types of vdW corrections of DFT-D2 and DFT-D3 were used for comparison \cite{D2-JCC-2006,D3-JCP-2010}. The convergence criterion of energy was set to $10^{-6}$ eV, and the criterion of Hellman-Feynman forces during the structural relaxation was $0.005$ eV/\AA.
	
The phonon spectrum calculation was used by the finite differences method \cite{Finite-differences-PRB-2002,Finite-difference-PRB-2005}. The polarization was calculated by the standard Berry Phase method \cite{Polarization-PRB-1993,Resta-RevModPhys-1994}. In order to describe the correlated $d$ electrons, the Hubbard $U_{\rm eff}$ with the Dudarev approximation was applied \cite{Dudarev1998-PRB}. In addition, the linear response ansatz was used to eastimate the appropriate $U_{\rm eff}$ \cite{Cococcioni-2005-PRB}.

\section{Results and Discussion}
\subsection{Crystalline \& electronic structures}
\begin{figure}
	\centering
	\includegraphics[width=0.47\textwidth]{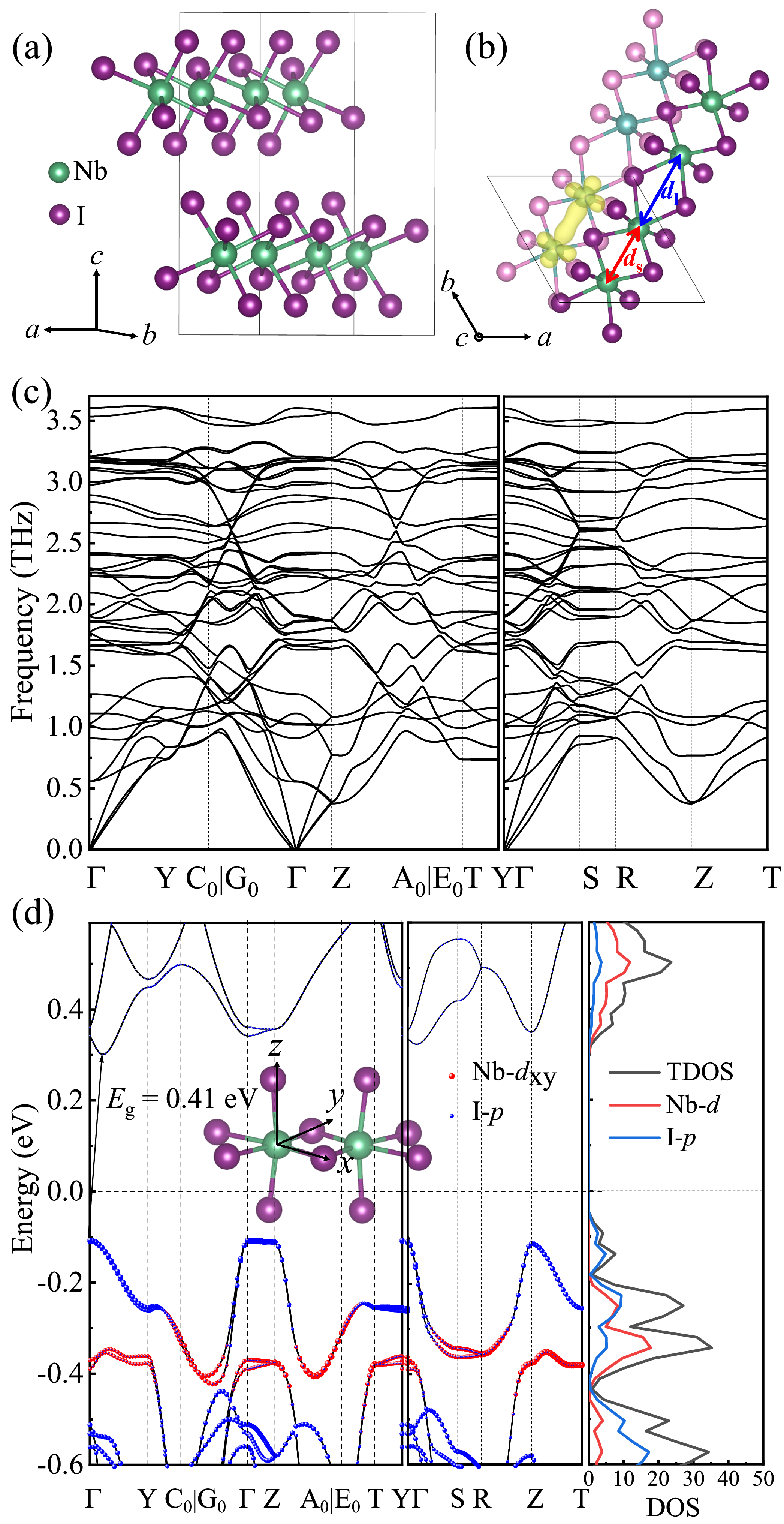}
	\caption{(a) Side view of NbI$_4$ bulk. Each unit cell (u.c.) contains two chains coupled by vdW force. (b) View of NbI$_4$ bulk from the $c$-axis. The structural dimerization of Nb ions can be indicated by the longer $d_l$ and shorter $d_s$. Correspondingly, the electron occupancy forms a molecular orbital between the shorter Nb-Nb pair, as indicated by the yellow shadow. (c) The phonon spectrum. (d) The electronic band structure and corresponding density of states (DOS). The red and bule curves indicate the contributions from Nb's $d$ and I's $p$ orbital respectively. Inset: the Cartesian coordinates used for orbital decomposition.}
	\label{fig1}
\end{figure}
Figure~\ref{fig1}(a-b) shows the structures of NbI$_4$ bulk in the primitive cell with the space group $Cmc2_1$ (No. $36$). The transformation matrix between the conventional (original experimental) cell [labeled as ($a$, $b$, $c$)] and the primitive cell [labeled as ($a'$, $b'$, $c'$)] with the space group $Cmc2_1$ is $(a, b, c)^{T}=\begin{pmatrix}
		1 & 1 & 0 \\
		-1 & 1 & 0 \\
		0 & 0 & 1
\end{pmatrix}(a', b', c')^{T}$, as depicted in Fig.~S1 \cite{sm}. The lattice constants optimized by DFT+D2 vdW correction with pure GGA are $a'$=$b'$=$7.648$ \AA\ and $c'$=$13.927$ \AA, which agree best with the experimental values $a'$=$b'$=$7.646$ \AA\ and $c'$=$13.930$ \AA\ \cite{dahl1962-AC}, as compared in Table~\ref{table1}. Thus, in the following calculation, the D2 correction will be used by default. In the primitive cell, there are two NbI$_4$ chains, and these two chains are assembled by vdW force.

Within each chain, every Nb ion is caged within the I$_6$ octahedron and the neighbor octahedra are edge-sharing. Interestingly, two adjacent Nb ions are dimerized, with a significant disproportion of nearest-neighbor Nb-Nb distances: the longer one $d_l$=$4.450$ \AA\ and the shorter one $d_s$=$3.316$ \AA\, as depicted in Fig.~\ref{fig1}(b). The phonon spectrum is calculated as shown in Fig.~\ref{fig1}(c), and there are no imaginary frequencies in the whole Brillouin zone, implying its dynamic stability. 
	
\begin{table}
\caption{The lattice constants of NbI$_4$ bulk from DFT calculations and experiment (Exp), which are in units of \AA. $\alpha$/$\beta$/$\gamma$ indicate the crystallographic axis angles, in unit of $\degree$. D2 and D3 indicate different vdW corrections in DFT calculations. The original experimental data are for an orthorhombic axes ($a$=$a'$, $b$=$\sqrt{3}b'$, $c$=$c'$). It is clear the DFT-D2 calculation leads to be best description of structure.}
\begin{tabular*}{0.47\textwidth}{@{\extracolsep{\fill}}ccccccc}
\hline \hline
& $a'$ & $b'$ & $c'$ & $\alpha$ & $\beta$ & $\gamma$  \\
\hline
GGA & 8.159 & 8.159 & 15.265 & 90 & 90 & 122.90\\
GGA+D2 & 7.648 & 7.648 & 13.927 & 90 & 90 & 118.98\\
GGA+D3 & 7.721 & 7.721 & 14.196 & 90 & 90 & 119.71\\
Exp~\cite{dahl1962-AC} & 7.646 & 7.646 & 13.930 & 90 & 90 & 119.79\\
\hline \hline  		
\end{tabular*}
\label{table1}
\end{table}
	
The calculated electronic structure indicates that NbI$_4$ is a semiconductor with an indirect band gap $0.41$ eV, as shown in Fig.~\ref{fig1}(d). The top valence bands are dominated by the the I's $5p$ and Nb's $4d_{xy}$ orbitals. For shorter Nb-Nb pairs, the overlap between two $4$d$^1$ electron orbitals are stronger, forming the bonding state (i.e. the $\sigma$ bond). The partial charge density for the top valence bands (within [-0.5, 0] eV) is calculated, which indicates the forming of a molecular orbital between the shorter Nb-Nb pairs, as shown in Fig.~\ref{fig1}(b). And the electronic band structure of NbI$_4$ without Nb-Nb dimerization is shown in the Fig..~S2 \cite{sm}, which indicates a metallic electronic structure. In this sense, the structural dimerization is a kind of Peierls transition driving the half-filling electron density. And this molecular orbital, i.e. the bonding state, is occupied by two electrons of Nb-Nb pair, lead to zero net magnetic moment. Since its point group ($mm2$) is polar, a polarization is estimated as $0.13$ $\mu$C/cm$^2$, along the $c$-axis.
	
\begin{figure}
\includegraphics[width=0.47\textwidth]{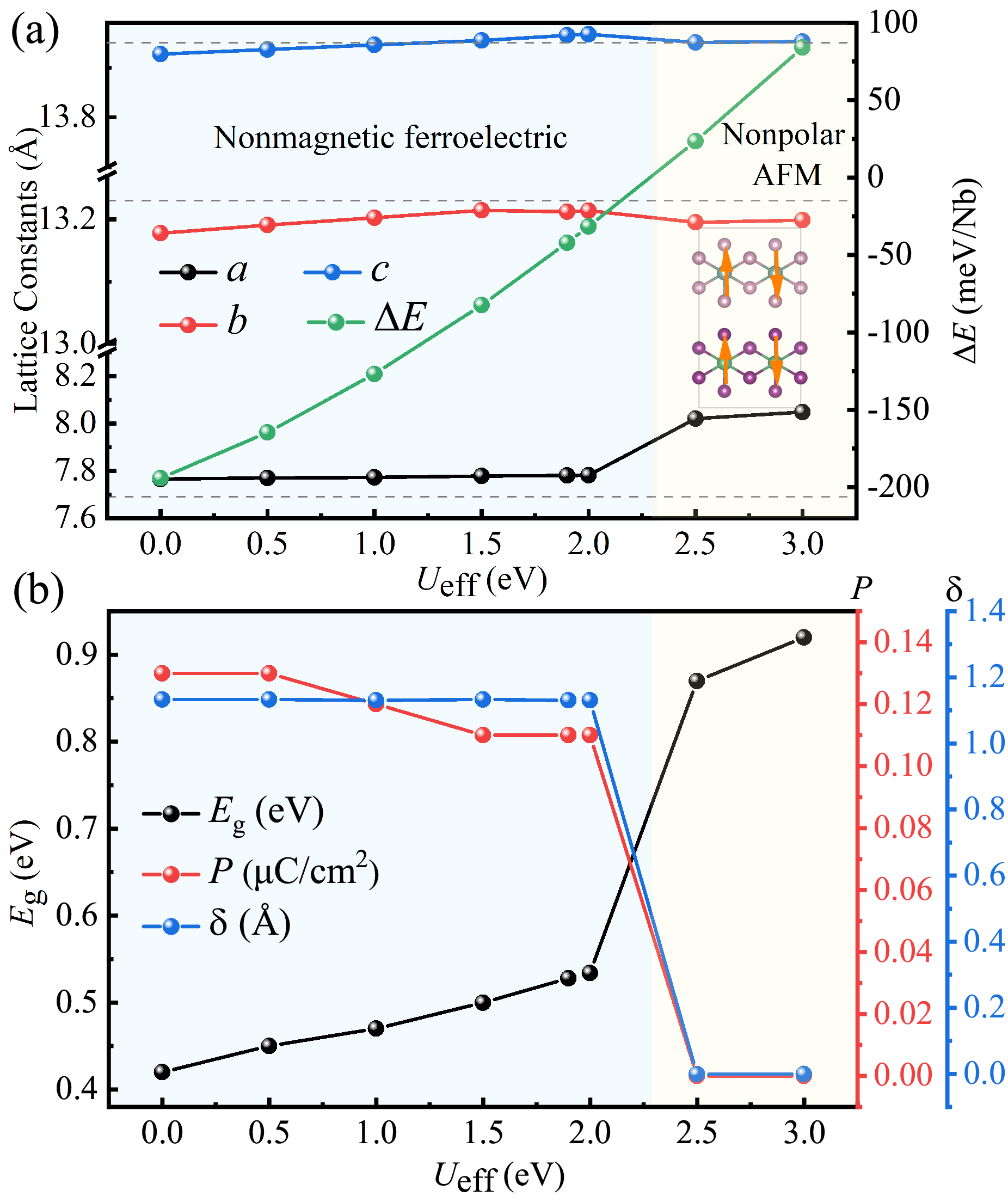}
\caption{(a) The lattice constants and energy difference ($\Delta E$) between the polar nonmagnetic (NM) state and nonpolar antiferromagnetic (AFM) state as a function of $U_{\rm eff}$. Inset: the magnetic order, which is N\'eel-type antiferromagnetic within each chain but ferromagnetic coupled between chains. However, the interchain exchange is much weaker than the intrachain one. Note, the basis vectors are uniformed to be the orthogonal ones ($a$, $b$, $c$). Here the lattice constant along the b-axis is doubled for the nonpolar antiferromagnetic state (Pbam). (b) Physical properties as a function of $U_{\rm eff}$, including the band gap ($E_{\rm g}$), polarization ($P$), and intensity of Peierls' distortion ($\delta$=$d_l$-$d_s$).}
\label{fig2}
\end{figure}

\begin{figure}
	\includegraphics[width=0.45\textwidth]{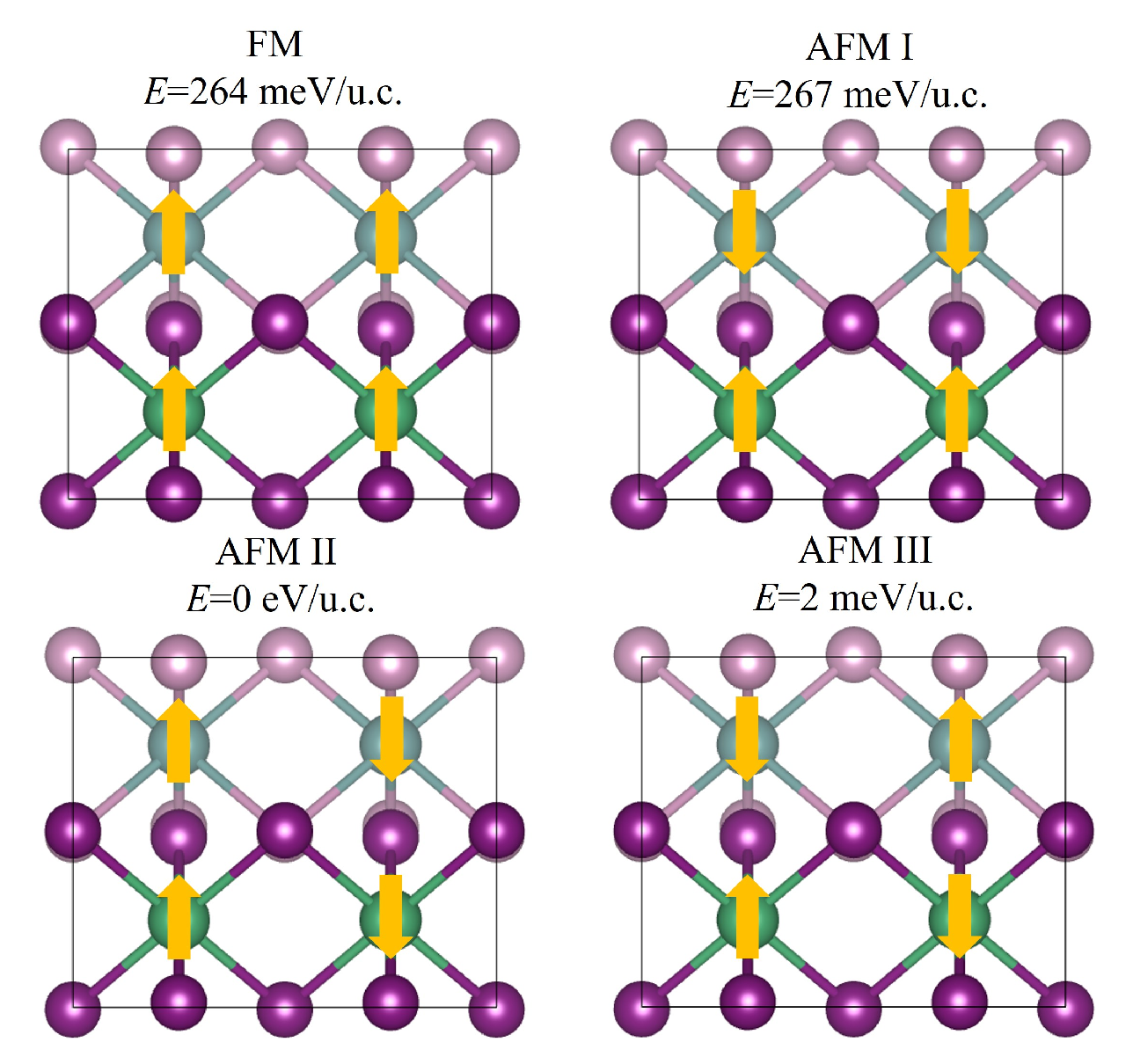}
	\caption{Possible magnetic orders and corresponding DFT energies. The ferromagnetic (FM) state is taken as the reference. The antiferromagnetic (AFM) II is the ground state, with N\'eel antiferromagnetic order in chain but FM coupling between chains.}
	\label{fig3}
\end{figure}

To account the possible Hubbard correction, the spin-polarized GGA+$U$ calculation is also performed. The physical properties such as lattice constants, band gap, and polarization are calculated as a function of $U_{\rm eff}$. Considering the $d^1$ fact, a simple N\'eel antiferromagnetic state (inset of Fig.~\ref{fig2}(a)) is also calculated for comparison. As shown in Fig.~\ref{fig2}(a), the nonmagnetic state is more energetic favorable, and its lattice constants are almost unaffected with increasing $U_{\rm eff}$ when $U_{\rm eff}\leq2$ eV. Since the Hubbard interaction tends to cause the Mott transition, namely to split the spin-up and spin-down channels to form local magnetic moments and a large Mott band gap. Here the Mott transition occurs once $U_{\rm eff}\geq2.5$ eV, so the magnetic state becomes more stable, with a expanding for lattice constant $a$ ($\sim3\%$) as a result of disappearance of Peierls transition. Accompanying this nonmagnetic-magnetic transition, a structural transition to $Pbam$ (No. $55$) occurs and the Peierls transition disappears, leading to the dissolution of molecular orbital. As a result, the local magnetic moment $\sim1$ $\mu_{\rm B}$/Nb appears, as expected for the $4d^1$ configuration. Consequently, the system changes from a band insulator to a Mott insulator with a larger band gap, as shown in Fig.~\ref{fig2}(b). Since the point group of magnetic state ($mmm$) is nonpolar, the polarization also disappears. Four possible magnetic orders are considered, with ferromagnetic/antiferromagnetic orders within/between chains as depicted in Fig.~\ref{fig3}. The DFT energies indicate that the N\'eel antiferromagnetic order is stable in each chain, while the interchain coupling is weakly ferromagnetic (AFM II). And its phonon spectrum is also dynamic stable, as shown Fig.~S3 \cite{sm}. In addition, spin-orbital coupling (SOC) effect is also taken into consideration to double-check the physical properties as depicted in Fig.~S4 \cite{sm}.

In short, the NbI$_4$ bulk can be nonmagnetic ferroelectric or nonpolar antiferromagnetic, depending on the Hubbard $U_{\rm eff}$. This behavior seems to be similar to the recently proposed idea of ``alterferroicity'' \cite{wang2023-PNAS}, if its real Hubbard $U_{\rm eff}$ can be tuned across the phase transition boundary. To test this idea, the effective Hubbard parameter $U_{\rm eff}$ is estimated using the linear response ansatz method \cite{Cococcioni-2005-PRB}, which leads to $1.9$ eV. Therefore, the ground state of NbI$_4$ bulk should be stay in the nonmagnetic ferroelectric side but indeed close to the transition boundary. The DFT results at $U_{\rm eff}=1.9$ eV agree well with the experimental structure: a polar structure and very precise lattice constants (the error bar is less than $1$\%). Therefore, all following calculations are performed with $U_{\rm eff}$=$1.9$ eV.
	
\subsection{Sliding ferroelectricity}
As aforementioned, the space group of NbI$_4$ bulk calculated by GGA+D2+$U$ ($U_{\rm eff}$ = $1.9$ eV) is $Cmc2_1$ (No. 36), which belongs to the polar point group $mm2$. Using the standard Berry phase calculation, the polarization is estimated as $0.11$ $\mu$C/cm$^2$, pointing along the $c$ axis, which is comparable to other typical sliding ferroelectrics, e.g., $0.03$ $\mu$C/cm$^2$ in WTe$_2$ bilayer and $0.68$ $\mu$C/cm$^2$ in $h$-BN bilayer \cite{Yang2018-JPCL,Liu-Nanoscale-2019,li-ACS-Nano-2017}.
	
\begin{figure}
\includegraphics[width=0.47\textwidth]{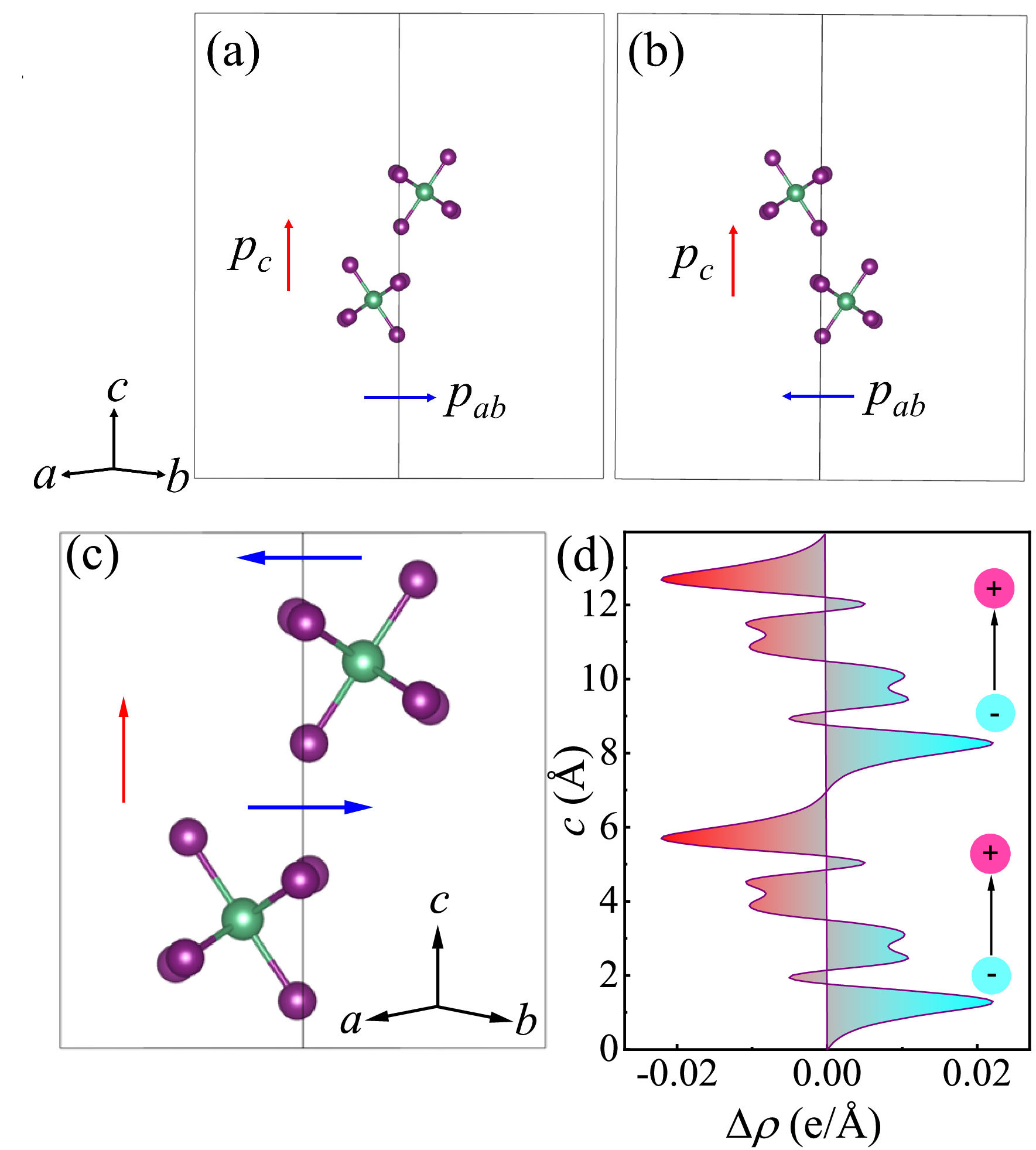}
\caption{Origin of interchain sliding ferroelectricity in NbI$_4$ bulk. (a-b) Isolated two chains with different binding modes: (a) A-B binding \textit{vs} (b) B-A binding. (c) Top view of chains in bulk. The induced dipoles are perpendicular to the chains. However, its component in the $ab$ plane (bule) is canceled in bulk, while only the component along the $c$-axis (red) gives raise to the net polarization. (d) The planar-averaged differential electron density $\Delta\rho$ between the $+P$ and $-P$ states along the $c$-axis. The $c$-position in (c) and (d) are one-to-one corresponding.}
\label{fig4}
\end{figure}
	
Although each chain of NbI$_4$ is non-polar, the specific binding mode of chains results in the polar structure, i.e. the interchain sliding. According to the empirical rule in 2D sliding ferroelectrics, the induced polarization should perpendicular to the sliding direction, i.e., along the out-of-plane axis. However, for the quasi-one-dimensional chains, the sliding is along the chain direction, and thus there are two perpendicular directions. As depicted in Fig.~\ref{fig4}(a-b), our calculation on isolated two chains indeed find two perpendicular components ($p_c$ \& $p_{ab}$) of dipole for each pair of chains. For isolated two chains, the calculated dipole along ($a$, $b$, $c$) is ($-0.0044$, $0.0044$, $-0.0682$) e\AA\ for the A-B binding mode, but ($0.0044$, $-0.0044$, $-0.0682$) e\AA\ for the B-A binding mode. Interestingly, the $p_c$'s are parallel between nearest-neighbor pairs. In contrast, $p_{ab}$'s are antiparallel and cancelled between nearest-neighbor pairs along the $c$-axis. As summarized in Fig.~\ref{fig4}(c), the induced polarization is only along the $c$-axis in its vdW bulk.

To trace the microscopic origin of dipoles, the planar-averaged differential electron density is calculated along the $c$ axis, defined as $\Delta\rho=\iint[\rho(+P)-\rho(-P)]dxdy$ between the opposite polarization states. As shown in Fig.~\ref{fig4}(d), the bias of electron cloud can be unambiguously visualized. As expected, the neighboring I ions between chains contibute most to the induced dipole.
	
\begin{figure}
\includegraphics[width=0.47\textwidth]{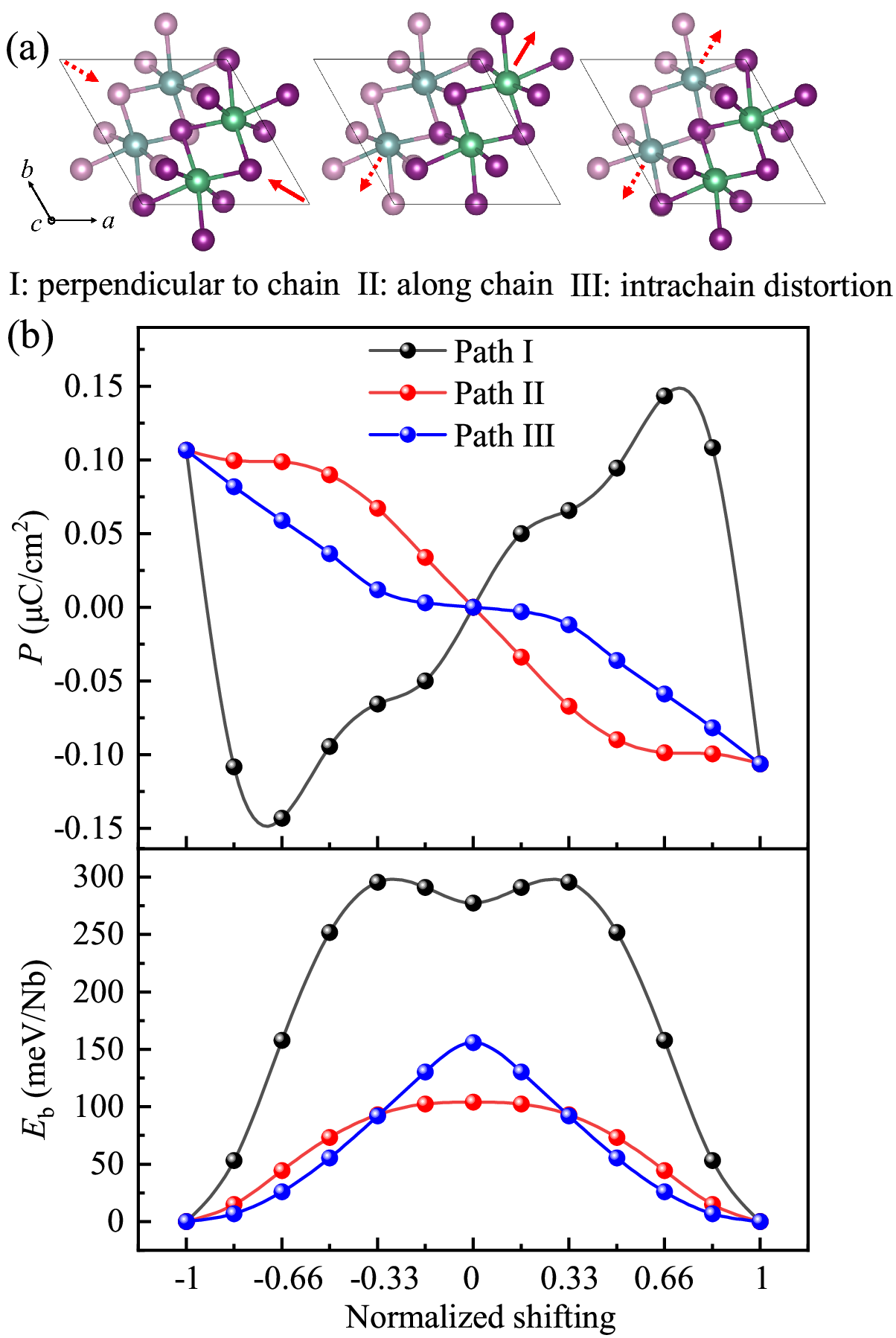}
\caption{(a) Three possible switching paths of polarization ($P$) in NbI$_4$ bulk. (b) Corresponding polarizations and energy barriers for these three switching paths. Horizontal axis: the interchain sliding and intrachain distortion are normalized.}
\label{fig5}
\end{figure}

The quasi-one-dimensional characteristic of NbI$_4$ provides more degrees of freedom regarding its sliding modes. Specifically, the slidings along the direction of chain and perpendicular to chain are shown in Fig.~\ref{fig5}(a), which can reverse the polarization. Besides, the polarization can also be reversed via the intrachain ion displacements, namely by altering the phase of Peierls dimerization in half of chains. These three  ferroelectric switching paths (labeled as I, II, and III) are inverstigated via linear interpolation. 

As shown in Fig.~\ref{fig5}(b), the polarizations and energy barriers of these three paths as a function of normalized shift are quite different. The path I, with interchain rigid sliding perpendicular to chains, has the highest energy barrier $\sim295$ meV/f.u.. Correspondingly, the change of polarization is not monotonic, but in an exotic trajectory. The path II, with interchain rigid sliding along the chain direction, has the lowest energy barrier $\sim100$ meV/f.u.. The path III, by changing the phase of Peierls dimerization in one chain, has a middle energy barrier $\sim156$ meV/u.c.. For both paths II and III, the polarization changes normally as a function of normalized shifting: decreases to zero first and then increases in the opposite direction. With the lowest energy barrier, the path II may be the most possible one in real switching process. In fact, the theoretical paths are all presumptive, which other possible switching modes (including those hybrid ones) can not be excluded in real cases.
	
\subsection{Negative piezoelectricity \& polar metal.}
Piezoelectricity is an important physical property of ferroelectric materials. In vdW ferroelectrics, negative piezoelectricity often appears as an inborn characteristic of low-dimensional materials \cite{you2019origin,Qi-PRL-2021}. However, previous studies on negative piezoelectricity were mostly focused on 2D materials and the investigation of piezoelectricity in quasi-one-dimensional sliding ferroelectrics is lacking. This work just fills the gap.

In order to investigate the piezoelectric property of NbI$_4$ bulk, the elastic matrix and the piezoelectric tensor matrix are calculated by energy-strain method and density functional perturbation theory (DFPT) method \cite{VASPKIT, DFPT-PRB-1997}, respectively. There are nine independent nonzero matrix elements for the space group $Cmc2_1$ in elastic matrix (in units of GPa) \cite{PhysRevB.90.224104}:
\begin{equation}
C=\begin{pmatrix}
73.404 & 13.015 & 11.936 & 0 & 0 & 0 \\
13.015 & 42.489 & 8.819 & 0 & 0 & 0 \\
11.936 & 8.819 & 45.130 & 0 & 0 & 0 \\
0 & 0 & 0 & 9.354 & 0 & 0 \\
0 & 0 & 0 & 0 & 12.062 & 0 \\
0 & 0 & 0 & 0 & 0 & 13.768
\end{pmatrix}.
\label{C}
\end{equation} 		
The piezoelectric tensor matrix has five independent nonzero matrix elements in units of C/m$^2$ for point group $mm$2 \cite{deJong2015}:	
\begin{equation}
e=\begin{pmatrix}
0 & 0 & 0 & 0 & -8.61 & 0 \\
0 & 0 & 0 & -0.77 & 0 & 0 \\
6.24 & 3.59 & -16.25 & 0 & 0 & 0
\end{pmatrix}.
\label{e}
\end{equation}	
Then the piezoelectric strain coefficient $d_{ij}$ can be calculated as: 
\begin{equation}
d_{ij}=\sum_{k=1}^6 e_{ik}C^{-1}_{kj},
\end{equation}
The piezoelectric strain coefficients $d_{ij}$ in units of pC/N are: 
\begin{equation}
d=\begin{pmatrix}
0 & 0 & 0 & 0 & -0.63 & 0 \\
0 & 0 & 0 & -0.06 & 0 & 0 \\
0.13 & 0.13 & -0.42 & 0 & 0 & 0
\end{pmatrix}.
\label{d}
\end{equation}
And the calculated $d_{33}$ is $-0.42$ pC/N, indicating a negative piezoelectricity in this quasi-one-dimensional sliding ferroelectric. It means that the polarization is enhanced when the lattice is shrinking along the polarization direction, or vice versa. The mechanism of the negative piezoelectricity of NbI$_4$ is similar to that of the interlayer-sliding ferroelectricity\cite{ding2021-PRM}, which is the joint contribution of dipole moment increase and volume decrease. In other words, the closer neighboring NbI$_4$ chains, the larger interchain sliding and the stronger dipole moment within smaller volume. Consequently, a larger polarization perpendicular to the chains is induced.

\begin{figure}
\includegraphics[width=0.48\textwidth]{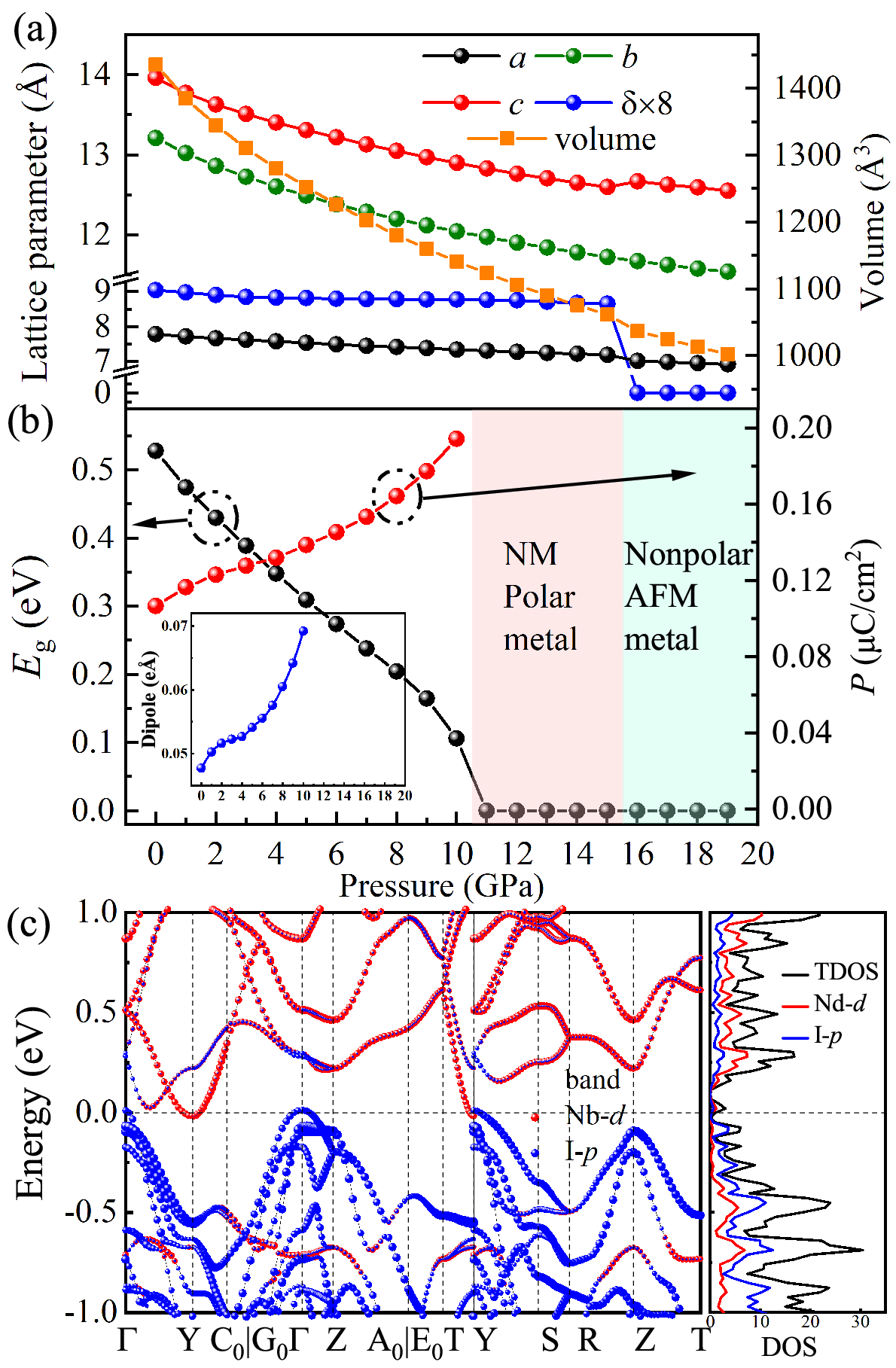}
\caption{(a) The lattice parameters including lattice constants and eight times Peierls dimerization (characterized by $\delta \times 8$) as a function of hydrostatic pressure. Note, the basis vectors are uniformed to be the orthogonal ones ($a$, $b$, $c$). The unit cell of $Cmc$2$_1$ one is doubling along the $b$-axis comparing with the $Pbam$ one. Thus, the corresponding lattice constant for $Pbam$ phase is $2b$. (b) Band gap and polarization as a function of hydrostatic pressure. Inset: the dipole of one unit cell as a function of pressure in the ferroelectric region. (c) The electron band structure and density of states under the pressure of $11$ GPa (i.e. in the polar metal region).}
\label{fig6}
\end{figure}
	
There are two competing phases (polar $\it{vs}$ magnetic) in the NbI$_4$ bulk, whose volumes are different. Naturally, the pressure is an effective strategy to regulate phase transitions. Finally, the effect of hydrostatic pressure is investigated. The lattice constants and Peierls dimerization are depicted in Fig.~\ref{fig6}(a)  as a function of hydrostatic pressure. As expected, the lattice constants decrease significantly. For example, at $9$ GPa, $a$, $b$ and $c$ shrink for $5.1\%$, $8.3\%$ and $7.1\%$ respectively and the volume is reduced for $19.2\%$. For comparison, the Peierls dimerization $\delta$ only decreases for $3.0\%$, due to the stiffness of intrachain.

In consistent with the negative piezoelectricity, the polarization is promoted for $82\%$ by hydrostatic pressure up to $10$ GPa, as shown in Fig.~\ref{fig6}(b). The physical origin is not only the reduction in volume but also the enhanced dipole (Inset of Fig.~\ref{fig6}(b)) of NbI$_4$ chains, which is also similar to the interlayer-sliding ferroelectricity \cite{ding2021-PRM}.

Under expectation, the band gap decreases with hydrostatic pressure, as shown in Fig.~\ref{fig6}(b). Beyond $10$ GPa, the band gap of NbI$_4$ bulk close while its polar structure remains. In other word, it becomes a polar metal. The typical electron band structure is shown in Fig.~\ref{fig6}(c). Such pressure induced insulator-metal transition belongs to the Bloch-Wilson transition, which also widely exists in pressurized superconductors~\cite{Imada-RevModPhys-1998}. With further increasing pressure ($\geq16$ GPa), the N\'eel-type antiferromagnetic state appears with suppressed Peierls dimerization based on the enthalpy difference between nomagnetic polar and nonpolar antiferromagnetic states (Fig.~S5 \cite{sm}). 

\section{Conclusion}	
In summary, based on the first-principles calculations, the structural and electronic properties of NbI$_4$ vdW bulk have been studied systematically. The structural Peierls dimerization within its one-dimensional chain, quenches the local magnetic moment of $4d^1$ electron. Binding between tow-chains induces the charge redistribution and further leads to the polarization, which can be switched by the interchain sliding. Besides, the negative piezoelectricity effect is predicted. In addition, the polarization can be enlarged for $82\%$ by hydrostatic pressure up to $10$ GPa, beyond which NbI$_4$ becomes a polar metal. With further increasing pressure ($\geq16$ GPa), the N\'eel-type antiferromagnetic state appears with suppressed Peierls dimerization and polarization.
	
\begin{acknowledgments}
This work was supported by the National Natural Science Foundation of China (Grant Nos. 12274069 and 12325401) and Jiangsu Funding Program for Excellent Postdoctoral Talent under Grant Number 2024ZB001. We thank the Big Data Center of Southeast University for providing the facility support on the numerical calculations.
\end{acknowledgments}	
\bibliography{apssamp}
\end{document}